\documentclass[journal,10pt]{IEEEtran}
\pdfoutput=1

\newlength{\figwidth}
\usepackage{color}
\definecolor{links}{rgb}{0.7,0,0}   % red
\definecolor{urls}{rgb}{0,0,0.8}    % blue
\definecolor{cites}{rgb}{0,0,0.8}   % blue

\usepackage[colorlinks,hyperindex,linkcolor=links,citecolor=cites,urlcolor=urls]{hyperref} % generates colored links in pdf file
 
\usepackage[nosort]{cite}        
\usepackage{url} 
\usepackage[intlimits]{amsmath}
\usepackage{bbm}
\usepackage{graphicx}
\usepackage{paralist}
\usepackage{fancyref}
\usepackage[stretch=16,shrink=16,step=4]{microtype}
\usepackage{vmr-symbols-vecbold}
\usepackage{standard-macros}
\usepackage{siunitx}
\makeatletter
\def\@IEEEinterspaceratioM{0.265}
\def\@IEEEinterspaceMINratioM{0.1651}
\def\@IEEEinterspaceMAXratioM{0.38}

\def\@IEEEinterspaceratioB{0.31}
\def\@IEEEinterspaceMINratioB{0.19}
\def\@IEEEinterspaceMAXratioB{0.38}
\@IEEEtunefonts
\makeatother
\safemath{\maxrate}{R^{*}}
\safemath{\txant}{m\sub{t}} %number of transmit antennas
\safemath{\rxant}{m\sub{r}} %number of receive antennas
\safemath{\cohtime}{n\sub{c}} %time-frequency coherence length (in symbol times)
\safemath{\tfdiv}{\ell}  %amount of time-frequency diversity available (number of independent fading realizations 
\begin{document}

\IEEEoverridecommandlockouts
% DRAFT
% to include revision information into the resulting PDF
%%%%\svnInfo $Id: dbs_it07.tex 2697 2009-08-04 10:43:20Z gdurisi $ 
%% add a PDFinfo field to store metadata in the output PDF

%\title{Towards Massive M2M and Low-Latency Wireless:  \\ The Art of Sending Short Packets}
%
\title{Towards Massive, Ultra-Reliable, and Low-Latency Wireless Communication with Short Packets}
\author{Giuseppe~Durisi,~\IEEEmembership{Senior Member,~IEEE,}
Tobias~Koch,~\IEEEmembership{Member,~IEEE,}
Petar~Popovski,~\IEEEmembership{Fellow,~IEEE}
\thanks{The work of G.~Durisi has been in part supported by the Swedish Research Council under Grant 2012-4571. The work of T.~Koch has been supported in part by the European Community's Seventh Framework Programme FP7/2007-2013 under Grant 333680, in part by the Ministerio de Econom\'ia y Competitividad of Spain under Grants TEC2013-41718-R, RYC-2014-16332, and TEC2015-69648-REDC, and in part by the Comunidad de Madrid under Grant S2013/ICE-2845. The work of P. Popovski has been in part supported by the European
Research Council (ERC Consolidator Grant Nr. 648382 WILLOW) within the Horizon 2020 Program.
The simulations were performed in part on resources at Chalmers Centre for Computational Science and Engineering (C3SE) provided by the Swedish National Infrastructure for Computing (SNIC). }
\thanks{G. Durisi is with the Department of Signals and Systems, Chalmers University of Technology, 41296,  Gothenburg, Sweden (e-mail: durisi@chalmers.se).}%
\thanks{T. Koch is with the Signal Theory and Communications Department, Universidad Carlos III de Madrid, 28911, Legan\'{e}s, Spain and also with the Gregorio Mara\~n\'on Health Research Institute, Madrid, Spain (e-mail: koch@tsc.uc3m.es).}
% \thanks{Y. Polyanskiy is with the Department of Electrical Engineering and Computer
% Science, MIT, Cambridge, MA, 02139 USA (e-mail: yp@mit.edu).}
\thanks{P. Popovski is with the Department of Electronic Systems, Aalborg University,
9220 Aalborg, Denmark (e-mail: petarp@es.aau.dk).}
\thanks{The authors are listed in alphabetical order.}

}
\maketitle

%%%%%%%%%%%%%%%%
 \begin{abstract}
Most of the recent advances in the design of high-speed wireless systems are based on information-theoretic principles that demonstrate how to efficiently transmit long data packets.
However, the upcoming wireless systems, notably the 5G system, will need to support novel traffic types that use short packets.
For example, short packets represent the most common form of traffic generated by  sensors and other devices involved in Machine-to-Machine (M2M) communications.
Furthermore, there are emerging applications in which small packets are expected to carry critical information that should be received with low latency and ultra-high reliability.

Current wireless systems are not designed to support short-packet transmissions. 
For example, the design of current systems relies on the assumption that the metadata (control information) is of negligible size compared to the actual information payload.
Hence, transmitting metadata using heuristic methods does not affect the overall system performance.
However, when the packets are short, metadata may be of the same size as the payload, and the conventional methods to transmit it may be highly suboptimal.

In this article, we review recent advances in information theory, which provide the theoretical  principles that govern the transmission of short packets. We then apply these principles to three exemplary scenarios (the two-way channel, the downlink broadcast channel, and the uplink random access channel), thereby illustrating how the transmission of control information can be optimized when the packets are short.
The insights brought by these examples suggest that new principles are needed for the design of wireless protocols supporting short packets. These principles will have a direct impact on the system design.
\end{abstract}
% section punch_lines (end)

%%%%%%%%%%%%%%%%%%%%%%%%%%%%%%%%%%
\section{Introduction} % (fold)
\label{sec:introduction}
The vision of the Internet of Things promises to bring wireless connectivity to ``\dots anything that may benefit from being connected\dots''~\cite{dahlman14-06a}, ranging from tiny static sensors to vehicles and drones.
A successful implementation of this vision calls for a wireless communication system that is able to support a much larger number of connected devices, and that is able to fulfill much more stringent requirements on latency and reliability than what current standards can guarantee.
Among the various current research and standardization activities, the one aimed at the design of fifth generation (5G) wireless systems stands out as the largest globally orchestrated effort towards addressing these challenges.

{So far, each new generation of cellular systems has been mainly designed with the objective to provide a substantial gain in data rate over the previous generation. 
5G will depart from this scheme: its focus will not only be on enhanced broadband services and, hence, higher data rates.
This is because the vast majority of wireless connections in 5G will most likely  be originated by autonomous machines and devices rather than by the human-operated mobile terminals for which traditional broadband services are intended.
5G will address the specific needs of autonomous machines and devices by providing two novel wireless modes:  ultra-reliable communication (URC) and massive machine-to-machine communications (MM2M)~\cite{boccardi14-02a,osseiran14-05a,popovski14-10a}.}

URC refers to communication services where data packets are exchanged at moderately low throughput (e.g., $50$ Mbit/s) but with stringent requirements in terms of reliability (e.g., $99.999\%$) and latency (e.g., $4$ ms). 
Example of URC include reliable cloud connectivity, critical connections for industrial automation, and reliable wireless coordination among vehicles~\cite{popovski14-10a,johansson15-06a,yilmaz15-06a}. 

With MM2M one refers to the scenario where a massive number of devices (e.g., \SI{10000}\!) needs to be supported within a given area.
This is relevant for large-scale distributed cyber-physical systems (e.g., smart grid) or industrial control. 
Also in this case, the data packets are short (and often contain correlated measurements) and reliability must be high to cope with critical events.

%Low latency may also required in real-time applications, such as the one envisaged within the framework of the so-called ``tactile internet''~\cite{fettweis14-02a}, although differently from UCR and MM2M, these applications will mostly require high throughput as well.

The central challenge with these two  new wireless modes is the capability to support \emph{short packet} transmission. Indeed, short packets are the typical form of traffic generated by sensors and exchanged in machine-type communications. This requires a fundamentally different design approach than the one used in current high-data-rate systems, such as 4G LTE and WiFi.

It is appropriate at this point to formally define what is meant by short/long packets. 
The transmission of a packet is a process in which the information payload (data bits) is mapped into a continuous-time signal, which is then transmitted over the wireless channel. 
A continuous-time signal with approximate duration $T$ and approximate bandwidth $B$ can be described by $n \approx BT$ complex parameters. It is then natural to refer to $n$ as the packet length, i.e., the number of degrees of freedom (channel uses) that are required for the transmission of the information payload.

{A \emph{channel code} defines a map between the information payload and the signal transmitted over the $n$ channel uses. 
The task of a wireless receiver is to recover the information payload from a distorted and noisy version of the transmitted signal.
A fundamental result in information theory~\cite{shannon48-07a} tells us that when $n$ is large (long packets), there exist channel codes for which the information payload can be reconstructed with high probability (in a sense we shall make precise in Section~\ref{sec:anatomy_of_a_packet}). 
Intuitively, when $n$ is large both the thermal noise and the distortions introduced by the propagation channel are averaged out due to the law of large numbers.
However, when $n$ is small (short packets) such averaging cannot occur. 
}

\begin{figure}[t]
  \centering
    \includegraphics[width=7cm]{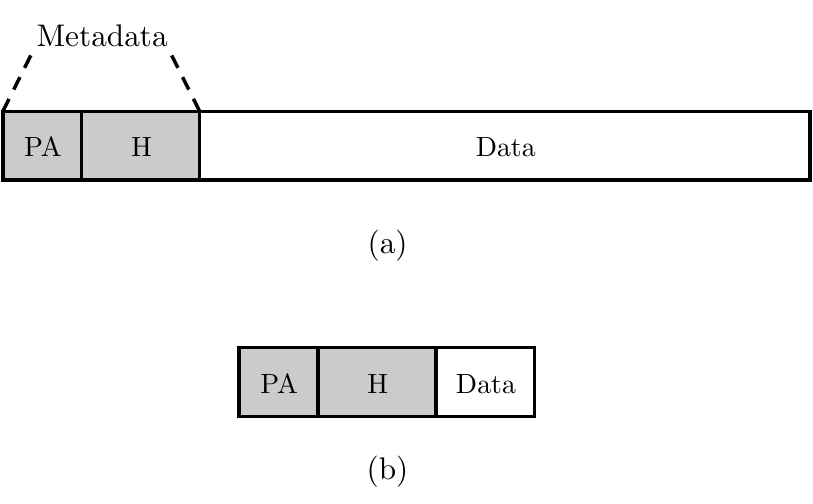}
  \caption{An example of a packet structure with data and metadata. In many wireless systems, metadata consists of preamble (PA) and header (H). 
  (a) Long data packet used in current wireless systems; (b) Short data packets needed to support novel 5G applications, such as URC and MM2M.}
  \label{fig:figs_data_metadata}
\end{figure}

Another defining element of long packets, besides the large number of channel uses, is the fact that  the payload contained in a packet is much larger than the control information (metadata) associated with the packet.
As a consequence, a highly suboptimal encoding of the metadata does not deteriorate the efficiency of the overall transmission, see Fig.~\ref{fig:figs_data_metadata}(a). 
On the contrary, when the packets are short, the metadata is no longer negligible in size compared to the payload, see Fig.~\ref{fig:figs_data_metadata}(b).

To summarize, in short-packet communications (i) classic information-theoretic results are not applicable because the law of large number cannot be put to work; (ii) the size of the metadata is comparable to the size of the payload and inefficient encoding of metadata significantly affects the overall efficiency of the transmission.

During the last few years, significant progress has been made within the information theory community to address the problem of transmitting short packets. Particularly for point-to-point scenarios, information theorists have gained some understanding of the theoretical principles governing short-packet transmission and possess metrics that allow them to assess their performance. In contrast, so far information theorists have mostly viewed the design of metadata as something outside their competence area. Consequently, the transmission of metadata has been largely left to heuristic approaches. 
In fact, practically all current protocols are based on a tacit assumption that the control information is perfectly reliable. 
A classic example is the proverbial ``one-bit acknowledgement'', which is always assumed to be perfectly received.

In this article, we present a comprehensive review of the theoretical principles that govern the transmission of short packets and present metrics that allow us to assess their performance. We then highlight the challenges that need to be addressed to optimally design URC and MM2M applications by means of three examples that illustrate how the tradeoffs brought by short-packet transmission affect protocol design.

{The paper is organized as follows. In Section~\ref{sec:anatomy_of_a_packet}, we describe the structure of a packet and review two classic information-theoretic metrics that are relevant for long packets: the \emph{ergodic capacity} and the \emph{outage capacity}. 
In Section~\ref{sec:rethinking_phy_performance_metrics},  we introduce a performance metric, the \emph{maximum coding rate} at finite packet length and finite packet error probability,  that is more relevant for the case of short packets. 
By focusing on the case of  additive white Gaussian noise (AWGN) channels and on the case of fading channels, we explain how to evaluate this quantity and discuss the engineering insights brought by it. 
In Section~\ref{sec:rethinking_the_mac_layer}, we illustrate through three example how to use the maximum coding rate performance metric to optimize the protocol design and the transmission of metadata in short-packet communications.
Concluding remarks are offered in Section~\ref{sec:conclusions}. }

%
%In this article, we will argue that supporting URC and MM2M will require a profound rethinking on how the physical (PHY) and medium access (MAC) layer of wireless communication systems are designed and on which metrics should be used to measure their performance.

% section introduction (end)

\section{Anatomy of a Packet} % (fold)
\label{sec:anatomy_of_a_packet}
Modern wireless systems transmit data in packets. 
Each transmitted packet over the air carries not only the information bits intended for the receiver but also additional bits that are needed for the correct functioning of the wireless protocols. 
{Such bits, which will be referred throughout as control information or metadata---in contrast to the actual \emph{data} to be transmitted---include packet initiation and termination, logical addresses, synchronization and security information, etc\dots
}

\begin{figure}[t]
  \centering
    \includegraphics[width=.49\textwidth]{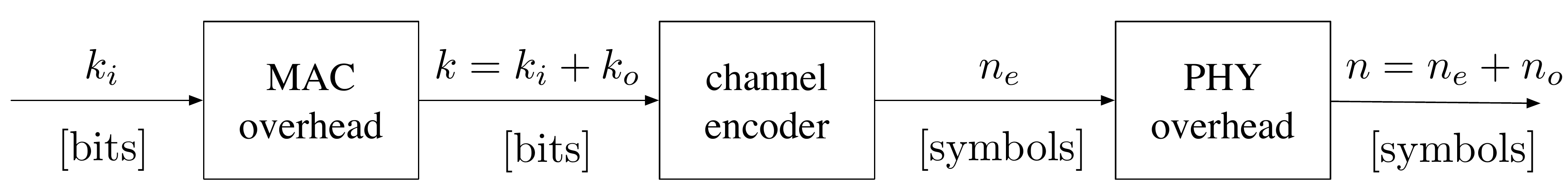}
  \caption{Block diagram that illustrates how a packet is created.}
  \label{fig:mac_phy}
\end{figure}

As illustrated in Fig.~\ref{fig:mac_phy}, a packet consists of $k$ payload bits, which are made up of $k_i$ information bits (information payload) and $k_o$ additional bits, containing metadata from the media-access-control (MAC) layer and higher layers. The payload bits are typically encoded into a block of $n_e$ data symbols (complex numbers) to increase reliability in packet transmission. Finally, $n_o$ additional symbols are added to enable packet detection, efficient synchronization (in time and frequency), or estimation of channel state information (CSI), which is needed by the receiver to compensate for the distortion of the transmitted signal introduced by the wireless channel. The total packet length $n$ is thus equal to $n_e+n_o$. With a slight abuse of notation, we shall refer to the additional $k_o$ bits and $n_o$ symbols as metadata.

The ratio $R = k_i/n$, i.e., the number of information bits per complex symbol (or, equivalently, the number of transmitted payload bits per second per unit bandwidth)  represents the net transmission rate and is a measure of the spectral efficiency of a communication system. 
In some wireless standards (such as LTE) specific physical/logical channels are reserved to carry exclusively metadata (control channels). This lowers further the spectral efficiency.

In most current wireless systems, we have that $k_i\gg k_o$ and that $n_e\gg n_o$, so the net transmission rate $R$ is roughly $k/n_e$. Consequently, the performance of such systems essentially depends on the efficiency of the channel code. 
Furthermore, $k_i$ (and hence also $n_e$) is typically large. It follows that information-theoretic metrics such as capacity~\cite{shannon48-07a} and outage capacity (also known as \emph{capacity-versus outage})~\cite{ozarow94-05a} are accurate, in spite of being defined for asymptotically large packet sizes. In summary, 
encoding the data payload using a good channel code allows for reliable transmission at rates close to the capacity of the underlying channel.

In order to facilitate the review of the relevant information-theoretic metrics, we shall need a reference communication channel. 
{A communication channel---the central part of a communication model---describes the relation between the input signal and the output signal over the available $n$ channel uses.
As mentioned in Section~\ref{sec:introduction}, each channel use corresponds to the transmission of a complex symbol.
Throughout most of the paper, we shall focus on the following channel model (and its multiple-antenna extension):
\begin{equation}
\label{eq:SingleAntennaGaussian}
Y_k = H_k X_k + W_k, \quad k\in\mathbb{N}.
\end{equation}
Here, $X_k$ denotes the complex symbol transmitted over the $k$th channel use, $Y_k$ is the corresponding channel output, $H_k$ is the channel coefficient that represents fading and other propagation phenomena and $W_k$ is the additive Gaussian noise, which we shall assume to be drawn from a stationary memoryless process.}

{If $H_k$ is taken equal to a deterministic constant $c$ independent of $k$ and known to transmitter and receiver, i.e., $H_k=c$ for all $k$, then~\eqref{eq:SingleAntennaGaussian} describes an AWGN channel. 
The AWGN channel is an example of an ergodic channel, that is, it exhibits an ergodic behavior over the duration of each packet (recall that the noise process $\{W_k\}$ is assumed stationary and memoryless). 
For such ergodic channels,  the relevant performance metric is the capacity $C$, defined as the largest rate $k/n_e$ for which the packet error probability can be made arbitrarily small by choosing $n_e$ sufficiently large. 
We shall treat the AWGN channel in more detail in Section~\ref{sec:awgn_channel}. 
Another example of an ergodic channel is the memoryless block fading channel, see Section~\ref{sec:a_finite_blocklength_theory_of_wireless_communications}. 
In this model, $H_k$, which can be thought of as a multiplicative noise, is assumed not known a priori by the transmitter/receiver and to vary according to a block-memoryless process. 
}

% In other words, $H_k$ is treated as a multiplicative noise, in addition to the additive Gaussian noise, but both can be combatted in a way that can guarantee a nonzero data rate with negligible error probability.}

{A completely different situation is the one in which the fading coefficient $H_k$ is random but does not depend on $k$, i.e., $H_k=H$. Hence, the fading coefficient stays constant over the packet duration~\cite[p.~2631]{biglieri98-10a},\cite[Sec.~5.4.1]{tse05a}. 
For this nonergodic channel, if $H$ can take arbitrarily small values, the error probability cannot be made small by choosing $n_e$ large. 
This is the case for most fading distributions, e.g., Rayleigh, Rician, and Nakagami. Indeed, when $|H|$ is small (deep fade), then the entire packet is lost, independently of its length. 
In such a nonergodic case, a relevant performance metric is the outage capacity $C_{\epsilon}$ (also known as capacity-versus-outage or $\epsilon$-capacity)---defined as the largest rate $k/n_e$ for which a packet error probability less than a fixed $\epsilon>0$ can be achieved by choosing $n_e$ sufficiently large.} 

We note that both capacity and outage capacity require that the codeword length $n_e$ (i.e., the packet size) and, hence, also the size of the data payload $k$ be large.
When the packets are short, the situation changes drastically. 
On the one hand, new information-theoretic performance metrics other than capacity or outage capacity are needed to capture the tension between reliability and throughput, as well as the cost incurred in exploiting time-frequency and spatial resources (PHY overhead). On the other hand, when the packets are short, the MAC overhead is significant and needs to be designed optimally, perhaps together with the data. 
We shall address the former issue in Section~\ref{sec:rethinking_phy_performance_metrics} and the latter issue in Section~\ref{sec:rethinking_the_mac_layer}.

\section{Rethinking PHY Performance Metrics} % (fold)
\label{sec:rethinking_phy_performance_metrics}

\subsection{Backing Off from the Infinite Blocklength Asymptotics} % (fold)
\label{sec:short_packets}
\begin{figure}[t]
  \centering
    \includegraphics[width=.49\textwidth]{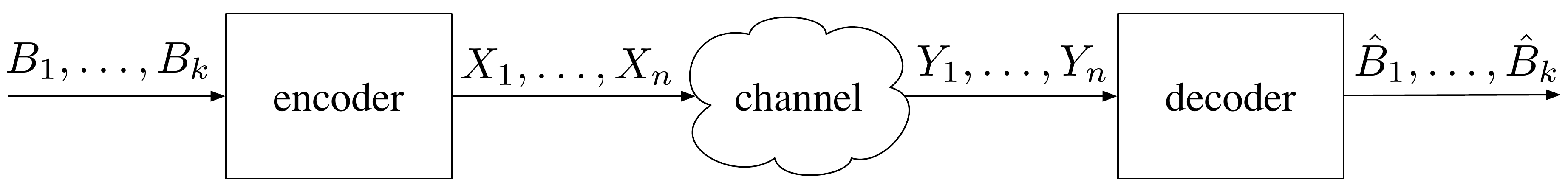}
  \caption{Information-theoretic description of a communication system. }
  \label{fig:IT_setup}
\end{figure}

In this section, we discuss information-theoretic performance metrics for short-packet wireless communications. We account for the metadata symbols required for the estimation of CSI, but ignore other issues such as packet detection or synchronization. As is common in information theory, we view the blocks \emph{channel encoder} and \emph{PHY overhead} in Fig.~\ref{fig:mac_phy} as one encoder block and consider the transmission of metadata symbols for channel estimation, such as \emph{pilot symbols}, as a possible encoding strategy. 
We note that the use of pilot symbols to estimate the channel is a widely adopted heuristic strategy that {may be} strictly suboptimal {in some cases}.

Mathematically, the encoder is modeled as a function $f_n$ that maps the $k$ information bits $B_1,\ldots,B_k$ to the sequence of symbols $X_1,\ldots,X_n$ to be transmitted over the channel; see Fig.~\ref{fig:IT_setup}. We shall refer to the number of transmitted symbols $n$ as the \emph{packet length} or \emph{blocklength} and to the sequence $X_1,\ldots,X_n$ as a \emph{codeword}. It is common to impose a power constraint $\rho$ on the transmitted symbols to account for restrictions on the transmit power, e.g., due to the devices' limited battery life or regulatory constraints. An often-used power constraint is the \emph{average power constraint}, under which the transmitted symbols must satisfy
\begin{equation}
\label{eq:average_power}
\frac{1}{n} \sum_{k=1}^n |X_k|^2 \leq \rho.
\end{equation}
The task of the decoder is to guess the information bits $B_1,\ldots,B_k$ from the $n$ channel outputs $Y_1,\ldots,Y_n$. The decoding procedure is modeled as a function $g_n$ that maps the channel outputs $Y_1,\ldots,Y_n$ to the estimates $\hat{B}_1,\ldots,\hat{B}_k$.

Let $P_{\text{e}}$ denote the \emph{packet error probability}, i.e., the probability that the decoder makes a wrong guess about the information bits $B_1,\ldots,B_k$. Note that $P_{\text{e}}$ does not only depend on the decoder $g_n$, but also on the encoder $f_n$. 

The rate $R$ of a communication system is defined as the fraction $k/n$ of information bits to the number of transmitted symbols. Ideally, we would like to design communication systems for which $R$ is as large as possible while, at the same time, the packet error probability $P_{\text{e}}$ is as small as possible. 
We denote by $\maxrate(n,\epsilon)$ {the \emph{maximum coding rate} at finite packet length $n$ and finite packet error probability $\epsilon$, i.e.,} the largest rate $k/n$ for which there exists an encoder/decoder pair $(f_n,g_n)$ of packet length $n$ whose packet error probability $P_{\text{e}}$ does not exceed $\epsilon$. 

Traditional information-theoretic metrics, such as \emph{capacity} \cite{shannon48-07a} and \emph{outage capacity} \cite{ozarow94-05a}, can be directly obtained from $\maxrate(n,\epsilon)$ by taking appropriate limits. 
Specifically, the outage capacity $C_{\epsilon}$ is defined as the largest rate $k/n$ such that, for every sufficiently large packet length $n$, there exists an encoder/decoder pair $(f_n,g_n)$ whose packet error probability does not exceed $\epsilon$. 
Thus, in contrast to $\maxrate(n,\epsilon)$, the definition of $C_{\epsilon}$ does not {involve} encoder/decoder pairs of a given {fixed} packet length $n$; instead, we consider encoder/decoder pairs whose packet lengths are large enough for the error probability to fall below $\epsilon$.
It follows that $C_{\epsilon}$ can be obtained from $\maxrate(n,\epsilon)$ via
\begin{IEEEeqnarray}{rCL}\label{eq:epsilon_capacity}
  %\IEEEeqnarraymulticol{3}{l}{...}
  % a & = & b +c
  C_{\epsilon} & = & \lim_{n\to \infty} \maxrate(n,\epsilon).
\end{IEEEeqnarray}
The capacity $C$ (in wireless communications also referred to as \emph{ergodic capacity}) is defined as the largest rate $k/n$ such that there exists an encoder/decoder pair $(f_n,g_n)$ whose packet error probability can be made arbitrarily small by choosing the packet length sufficiently large. Thus, in contrast to the definition of the outage capacity that demands a packet error probability smaller than some $\epsilon$, the definition of capacity is stronger in that it demands an arbitrarily small packet error probability. It follows that $C$ can be obtained from $C_{\epsilon}$ by letting $\epsilon$ tend to $0$:
\begin{IEEEeqnarray}{rCL}\label{eq:capacity}
  %\IEEEeqnarraymulticol{3}{l}{...}
  % a & = & b +c
  C & = & \lim_{\epsilon\to 0} C_{\epsilon} = \lim_{\epsilon\to 0} \lim_{n\to\infty} \maxrate(n,\epsilon).
\end{IEEEeqnarray}

Intuitively, the capacity characterizes the largest transmission rate at which reliable communication is feasible when there are no restrictions on the packet length. Likewise, the outage capacity characterizes the largest transmission rate at which communication with packet error probability not exceeding $\epsilon$ is feasible, again provided that there are no restrictions on the packet length.
It follows that both quantities are reasonable performance metrics for current wireless systems, where the packet size is typically large. However, assessing the performance of short packet communications requires a more refined analysis of $\maxrate(n,\epsilon)$. Unfortunately, the exact value of $\maxrate(n,\epsilon)$ is unknown even for channel models that are much simpler to analyze than the one encountered in wireless communications. Indeed, determining $\maxrate(n,\epsilon)$ is in general an NP-hard problem~\cite{costa10-06a}, and its complexity is conjectured to be doubly exponential in the packet length $n$.

Fortunately, during the last few years, significant progress has been made within the information theory community to address the problem of quantifying $\maxrate(n,\epsilon)$ and, hence, solve the long-standing problem of accounting for latency constraints in a satisfactory way.
Building upon Dobrushin's and Strassen's previous asymptotic results, Polyanskiy, Poor, and Verd\'u~\cite{polyanskiy10-05a} recently provided a unified approach to obtain tight bounds on $\maxrate(n,\epsilon)$.
They showed that for various channels with positive capacity $C$, the maximal coding rate $\maxrate(n,\epsilon)$ can be expressed as
\begin{IEEEeqnarray}{rCL}\label{eq:approximation}
  \maxrate(n,\epsilon) = C-\sqrt{\frac{V}{n}}Q^{-1}(\epsilon) + \mathcal{O}\lefto(\frac{\log n}{n}\right)
\end{IEEEeqnarray}
where $\mathcal{O}(\log n/n)$ comprises remainder terms of order $\log n/n$.
Here, $Q^{-1}(\cdot)$ denotes the inverse of the Gaussian $Q$ function and $V$ is the so-called \emph{channel dispersion}~\cite[Def.~1]{polyanskiy10-05a}.
The approximation~\eqref{eq:approximation} implies that to sustain the desired error probability $\epsilon$ for a given packet size $n$, one incurs a penalty on the rate (compared to the channel capacity) that is proportional to $1/\sqrt{n}$.

We next provide an interpretation for~\eqref{eq:approximation}.
The classic approach of approximating~$\maxrate(n,\epsilon)\approx C$ for large packet sizes and small packet error rates according to~\eqref{eq:capacity} allows one to model a communication channel as a ``bit pipe'' that delivers reliably $C$ bits per channel use.
This holds under the assumption that good channel codes are used.
The expansion provided in~\eqref{eq:approximation} suggests the following alternative model, which is more accurate when the packets are shorts:
A communication channel can be thought of as a bit pipe of randomly varying size.
Specifically, the size of the bit pipe behaves as a Gaussian random variable with mean $C$ and variance $V/n$.
Hence, $V$ is a measure of the channel dispersion. 
In this interpretation, the packet error probability $\epsilon$ is the probability that $\maxrate(n,\epsilon)$ is larger than the size of the bit pipe.

\subsection{AWGN Channel}
\label{sec:awgn_channel} 
\begin{figure}[t]
   \centering
     \includegraphics[width=.49\textwidth]{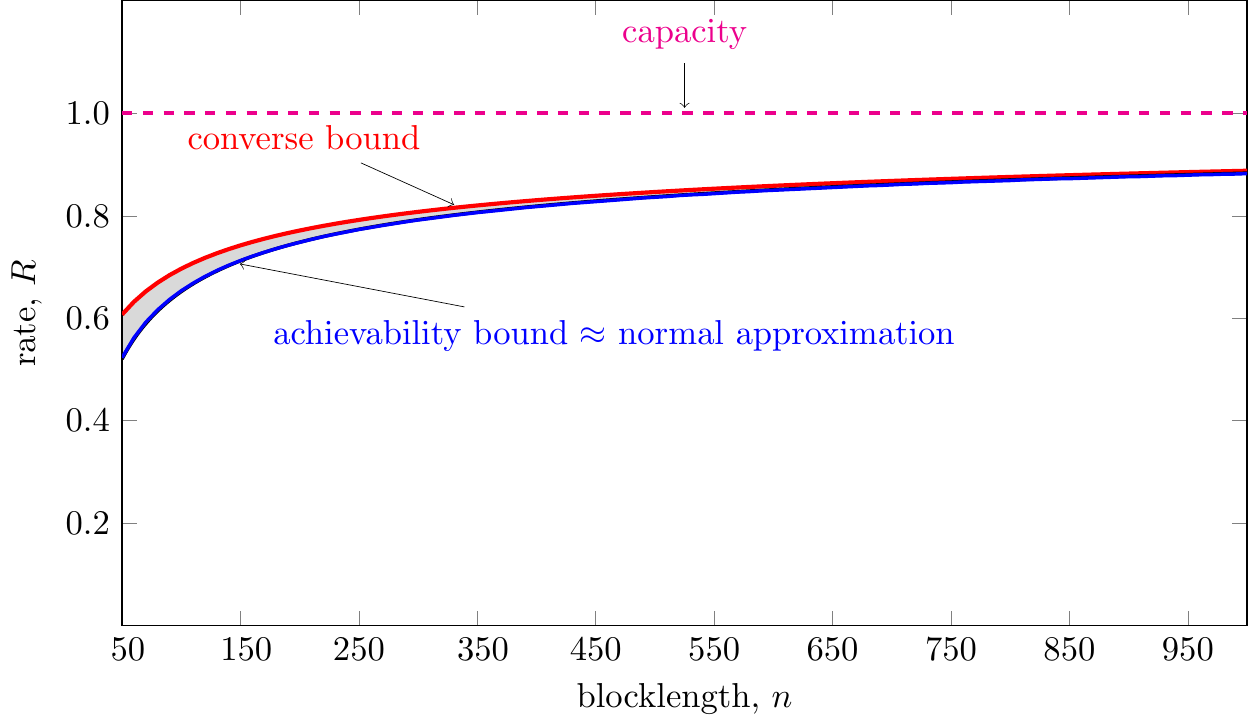}
   \caption{Upper bounds, lower bounds, and normal approximation on $\maxrate(n,\epsilon)$ for the AWGN channel with SNR $\rho=0$ dB. The packet error probability $\epsilon$ is $10^{-3}$.
   The upper bound is obtained using the metaconverse theorem~\cite[Th.~41]{polyanskiy10-05a}; the lower bound is the Shannon cone-packing bound~\cite{shannon59-a},\cite[Eq.~(41)]{polyanskiy10-05a}. The normal approximation is indistinguishable from the lower bound.}
   \label{fig:figs_awgn_plot}
 \end{figure}
Arguably, one of the best-understood channel models in the information theory literature is the average-power constrained AWGN channel.
Its canonical form can be obtained from~\eqref{eq:SingleAntennaGaussian} by setting $H_k=1$, which yields %
\begin{equation}
\label{eq:AWGN}
Y_k = X_k + W_k.
\end{equation}
Here, the inputs $\{X_k\}$ satisfy the average-power constraint \eqref{eq:average_power}.
When the additive noise has unit variance, the power constraint $\rho$ becomes equal to the signal-to-noise ratio (SNR).

For the AWGN channel, the capacity and the channel dispersion are given by \cite[Th.~54]{polyanskiy10-05a}\footnote{The capacity of the \emph{real-valued} AWGN channel has been obtained by Shannon \cite{shannon48-07a}. The channel dispersion of the \emph{real-valued} AWGN channel has been reported in~\cite[Eq.~(293)]{polyanskiy10-05a}. One obtains~\eqref{eq:C_Gauss} and \eqref{eq:V_Gauss} by noting that the transmission of a codeword of blocklength $n$ over the \emph{complex-valued} AWGN channel corresponds to the transmission of a codeword of blocklength $2n$ over the \emph{real-valued} AWGN channel with the same SNR.}
\begin{IEEEeqnarray}{rCL}
  C(\rho)&=& \log(1+\rho)\label{eq:C_Gauss}\\
  V(\rho)&=& \rho\frac{(2+\rho)}{(1+\rho)^2}(\log e)^2.\label{eq:V_Gauss}
\end{IEEEeqnarray}
It has been observed that a good approximation for $\maxrate(n,\epsilon)$ can be obtained by replacing the remainder terms on the {right-hand side} of \eqref{eq:approximation} by $(\log n)/(2n)$ \cite{polyanskiy10-05a,tan13-11a}. 
The resulting approximation, which is commonly referred to as \emph{normal approximation}, is plotted in Fig.~\ref{fig:figs_awgn_plot}, together with nonasymptotic upper and lower bounds on $\maxrate(n,\epsilon)$ (see~\cite{polyanskiy10-05a} for details).
 
As shown in the figure, the upper and lower bounds provide an accurate characterization of $\maxrate(n,\epsilon)$, which lies in the shaded region.
According to the bounds, to operate at $70\%$ of capacity with a packet error rate of $10^{-3}$, i.e., at $0.7$ bits/channel use, it is sufficient to use codes whose blocklength is between $110$ and $138$ channel uses. 
For the parameters considered in the figure, the normal approximation is indistinguishable from the achievability bound.
We also see that capacity is an inaccurate performance metric for packet sizes that are as short as the ones considered in the figure.

\subsection{Fading Channels} % (fold)
\begin{figure}[t]
  \centering
    \includegraphics[width=.25\textwidth]{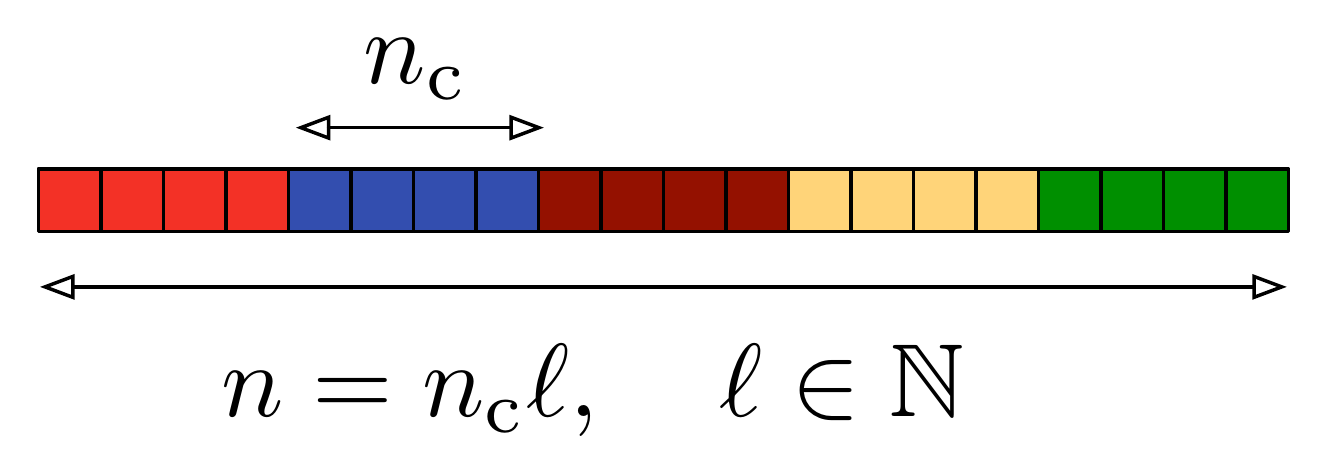}
  \caption{Block-fading model: the fading coefficient stays constant over $\cohtime$ channel uses (coherence interval) and then changes to an independent realization. Coding is performed over $\tfdiv$ coherence intervals (number of time-frequency diversity branches).}
  \label{fig:fbl-magazine_fbl-magazine-draft_figs_constant-block-fading}
\end{figure}

\label{sec:a_finite_blocklength_theory_of_wireless_communications}
We shall next discuss how to extend the results reported in Section~\ref{sec:awgn_channel} for the AWGN case to multiple-antenna fading channels.
Throughout, we shall focus on the \emph{memoryless block-fading model}~\cite{marzetta99-01a}, depicted in Fig.~\ref{fig:fbl-magazine_fbl-magazine-draft_figs_constant-block-fading}, according to which the fading coefficient stays constant for $\cohtime$ channel uses and then changes independently. In general, $\cohtime$ can be interpreted as the number of ``time-frequency slots" over which the channel does not change. We shall refer to each interval over which the fading coefficients do not change as a \emph{coherence interval}. 

The memoryless block-fading model is perhaps the simplest model to capture channel variations in wireless channels. 
Although inferior in accuracy to stationary channel models, where the channel varies continuously (see e.g.,~\cite{matz11-03a}), its simplicity enables analytical approaches that are currently out of reach for more sophisticated models.

For ease of notation, we shall write the symbols to be transmitted in each coherence interval in a $\cohtime\times\txant$ matrix whose entry at position $(i,j)$ corresponds to the $i$th symbol transmitted from antenna $j$. Likewise, we write the received symbols in a $\cohtime\times\rxant$ matrix. Within the $k$th coherence interval, the input-output relation of the block-fading channel with $\txant$ transmit and $\rxant$ receive antennas is given by
\begin{equation}
\label{eq:fading}
\mathbb{Y}_{k} = \mathbb{X}_{k} \mathbb{H}_{k} + \mathbb{W}_{k}.
\end{equation}
Here, $\mathbb{X}_{k}\in\mathbb{C}^{\cohtime\times \txant}$ and $\mathbb{Y}_{k}\in\mathbb{C}^{\cohtime\times \rxant}$ are the transmitted and received matrices, respectively; $\mathbb{H}_{k}\in\mathbb{C}^{\txant\times \rxant}$ denotes the fading matrix; $\mathbb{W}_{k} \in\mathbb{C}^{\cohtime\times \rxant}$ denotes the additive noise, which is assumed to have independent and identically distributed (\iid), zero-mean, unit-variance, complex Gaussian entries. For the sake of simplicity, we assume \emph{Rayleigh fading}, i.e., we assume that the fading matrix $\mathbb{H}_{k}$ has \iid, zero-mean, unit-variance, complex Gaussian entries. However, this assumption is not essential. In fact, most results presented in this paper were either originally derived for more general fading distributions or can be generalized with some effort.
For convenience, we shall assume that each codeword spans $\tfdiv$ coherence intervals, i.e., $n=\tfdiv\cohtime$.

We shall say that CSI is available at the transmitter, receiver, or both if the corresponding blocks have access to the realization of $\mathbb{H}_1,\ldots,\mathbb{H}_n$. In practice, CSI at the transmitter allows for transmission strategies that make use of the actual fading realization, thereby using the available transmit power more efficiently; CSI at the receiver facilitates the decoding task. Note that CSI at the receiver can be acquired by transmitting training sequences (so-called \emph{pilots}) that are used at the receiver to estimate the channel. CSI at the transmitter can, for example, be established by feeding channel estimates from the receiver back to the transmitter. However, the transmission of training sequences incurs a rate loss, sometimes referred to as \emph{channel-estimation overhead}. Likewise, the creation of a feedback link is associated with additional costs or overheads. Analyses relying on the assumption that CSI is available at the transmitter, receiver or both simply ignore these overheads. In this spirit, analyses that are based on the assumption that no CSI is available at the receiver do not assume that the receiver does not perform a channel estimation. On the contrary, they account for the overhead associated with the acquisition of CSI. For example, the transmission of training sequences can be viewed as a specific form of coding. Thus, by analyzing the fading channel \eqref{eq:fading} under the assumption that no CSI is available at the receiver, the rate loss incurred by the transmission of pilot symbols is automatically accounted for.

\subsubsection{Capacity-versus-outage at finite blocklength}
\label{sub:capacity-vs-outage}
We shall first discuss the case where the channel remains constant over the packet duration, i.e., $\tfdiv=1$. 
In this case, the fading channel is said to be \emph{quasi-static}, to reflect that the fading matrix is random but stays constant during the packet transmission.\footnote{In the information theory literature, the quasi-static channel model belongs to the class of \emph{composite channels} \cite{biglieri98-10a,effros10-07a}, also known as \emph{mixed channels}~\cite[Sec.~3.3]{han03-a}.} When communicating over quasi-static fading channels at
a given rate $R$, the realization of the random fading matrix $\mathbb{H}_k$ may be very small, in which case the decoder will not be able to guess the transmitted information bits correctly, no matter how large we choose the packet length $n$. In this case, the channel is said to be in \emph{outage}. For fading distributions for which the fading coefficient  can  be  arbitrarily small (which is, for example, the case for Rayleigh fading), the probability of an outage is  positive. Hence, the packet error probability is bounded away from zero for every positive rate $R>0$ and the capacity, defined as the largest rate for which reliable communication is feasible, is zero \cite{ozarow94-05a,biglieri98-10a}.

One may argue that the definition of capacity is too restrictive for quasi-static channels. Indeed, for sufficiently small (but positive) rates, the probability that the channel is in outage is typically small. Thus, while reliable communication cannot be guaranteed because there is always a chance that the channel is in outage, the probability that this happens is small. In other words, most of the time the channel is not in outage and reliable communication can be achieved by choosing a sufficiently large packet length. Capacity, however, is determined by those rare events where the channel is in outage. For applications where a positive packet error probability is acceptable, the outage capacity $C_{\epsilon}$ is arguably a more relevant performance metric than capacity, because it allows for outage events as long as they happen with probability less than $\epsilon$.

The outage capacity is often regarded as a performance metric for delay-constrained communication over slowly-varying fading channels (see, e.g., \cite{caire99-07a}). In fact, the assumption that the fading matrix stays constant during the packet transmission seems plausible only if the packet size is small. Nevertheless, the definition of outage capacity requires that the blocklength tends to infinity; cf.\ \eqref{eq:epsilon_capacity}. For example, for a single-antenna system, the outage probability as a function of the rate $R$ is given by \cite{telatar99-11a,caire99-07a,effros10-07a}
\begin{equation}
\label{eq:Pout_1}
P_{\text{out}}(R) = \Prob\lefto[\log\left(1+|H|^2 \rho\right)<R\right]
\end{equation}
and the outage capacity $C_{\epsilon}$ is the supremum of all rates $R$ satisfying $P_{\text{out}}(R)\leq \epsilon$, namely

\begin{equation}\label{eq:outage_cap_1}
  C_{\epsilon}=\sup\{R \sothat P_{\text{out}}(R)\leq \epsilon\}.
\end{equation}
The rationale behind this result is that, for every realization of the fading coefficient $H=h$, the quasi-static fading channel can be viewed as an AWGN channel with channel gain $h$, for which communication with arbitrarily small packet error probability is feasible if, and only if, $R<\log(1+|h|^2\rho)$, provided that the packet length is sufficiently large.\footnote{Indeed, the capacity of the AWGN channel with channel gain $h$ follows from \eqref{eq:C_Gauss} by changing the SNR from $\rho$ to $|h|^2\rho$.} However, it is \emph{prima facie} unclear whether the quantity $\log(1+|h|^2\rho)$ is meaningful when the packet size is small.

To better understand the relevance of the outage capacity for delay-constrained communication, a more refined analysis of $\maxrate(n,\epsilon)$ was presented in~\cite{yang14-07c}. 
It was shown that~\cite[Ths.~3 and~9]{yang14-07c}
\begin{equation}
\label{eq:approximation_QS}
\maxrate(n,\epsilon) = C_{\epsilon} + \mathcal{O}\lefto(\frac{\log n}{n}\right)
\end{equation}
irrespective of the number of transmit and receive antennas, and irrespective of whether CSI is available to transmitter, receiver, or both. Comparing \eqref{eq:approximation_QS} with \eqref{eq:approximation}, we observe that for the quasi-static fading case the channel dispersion is zero, i.e., the $1/\sqrt{n}$ rate penalty is absent. This suggests that $\maxrate(n,\epsilon)$ converges quickly to $C_{\epsilon}$ as $n$ tends to infinity, thereby indicating that the outage capacity is indeed a meaningful performance metric for delay-constrained communication over slowly-varying fading channels. Numerical examples that support this claim can be found in \cite[Sec.~VI]{yang14-07c}. 
Furthermore, a simple approximation for $\maxrate(n,\epsilon)$ is proposed in~\cite[Eqs.~(59) and~(95)]{yang14-07c}.
For the single-antenna case, this approximation can be written in the following form~\cite{yang14-07c,molavianJazi13-10}
\begin{IEEEeqnarray}{rCL}
  \epsilon\approx \Ex{}{Q\lefto(\frac{C(\snr\abs{H}^2)+{(\log n)}/({2n}) -\maxrate(n,\epsilon)}{\sqrt{V(\snr\abs{H}^2)/n}}\right)}. \IEEEeqnarraynumspace
\end{IEEEeqnarray}
Here, $C(\cdot)$ and $V(\cdot)$ are the functions defined in~\eqref{eq:C_Gauss} and~\eqref{eq:V_Gauss}, respectively.

The asymptotic expansion~\eqref{eq:approximation_QS} provides mathematical support to the observation reported by several researchers in the past that the outage probability describes accurately the performance over quasi-static fading channels of actual codes (see \cite{caire99-07a} and references therein). 
The intuition behind this result is that the dominant error event over quasi-static fading channels is that the channel is in a deep fade. Since the transmitted symbols experience all the same fading, it follows that coding is not helpful against deep fades in the quasi-static fading scenario, hence $R^*(n,\epsilon)$ is close to $C_{\epsilon}$ already for small blocklengths.

It has been observed that the outage capacity $C_{\epsilon}$ does not depend on whether CSI is available at the receiver \cite[p.~2632]{biglieri98-10a}, \cite[Ths.~3 and~9]{yang14-07c}. Intuitively, this is true because the fading matrix stays constant during the whole transmission, so it can be accurately estimated at the receiver through the transmission of $\sqrt{n}$ pilot symbols with no rate penalty as the packet length $n$ tends to infinity. This in turn implies that the outage capacity does not capture the channel-estimation overhead. Consequently, outage capacity is an inaccurate performance metric when the coherence interval $\cohtime$ is small.

\subsubsection{Tradeoff between diversity, multiplexing, and channel estimation}
When communicating over multiple-input multiple-output fading channels, a {crucial} question is whether the spatial degrees of freedom offered by the antennas should be used to lower the packet error probability for a given data rate (through the exploitation of \emph{spatial diversity}) or to increase the data rate for a given packet error probability (through the exploitation of \emph{spatial multiplexing}). These two effects cannot be harvested concurrently, but there exists a fundamental tradeoff between diversity and multiplexing. This tradeoff admits a particularly simple characterization in the high-SNR regime \cite{zheng03-05a}.

Specifically, Zheng and Tse \cite{zheng03-05a} defined the \emph{diversity-multiplexing tradeoff} as follows. Assume that $\tfdiv$ and $\cohtime$ are fixed. Further assume that the packet error probability vanishes with increasing $\rho$ as
\begin{equation}
\epsilon(\rho) = \rho^{-d\tfdiv}
\end{equation}
where $d\in\{1,\ldots,\txant\rxant\}$ is the so-called \emph{spatial diversity gain}. The \emph{multiplexing gain} $r(d)$ corresponding to the diversity gain $d$ is defined as
\begin{equation}
r(d) = \lim_{\rho\to\infty} \frac{\maxrate\bigl(n,\epsilon(\rho)\bigr)}{\log\rho}.
\end{equation}
For the case where CSI is available at the receiver and $\cohtime\geq \txant$, one can show that $r(d)$ is the piecewise linear function connecting the points \cite{zheng03-05a,elias06-01a}
\begin{equation}
\label{eq:r(d)_coh}
r\bigl((\txant-k)(\rxant-k)\bigr) = k, \quad k=0,\ldots,\min\{\txant,\rxant\}.
\end{equation}
Let $m^*=\min\{\txant,\rxant,\lfloor\cohtime/2\rfloor\}$, where $\lfloor a \rfloor$ denotes the largest integer that is not larger than $a$. For the case where no CSI is available at the receiver and $\cohtime\geq 2m^*+\rxant+1$, the multiplexing gain is given by \cite{zheng02-10a,zheng02-11a}
\begin{equation}
\label{eq:r(d)_noncoh}
r\bigl((\txant-k)(\rxant-k)\bigr) =\left(1-\frac{m^*}{\cohtime}\right) k.
\end{equation}
It is thus equal to \eqref{eq:r(d)_coh} multiplied by $(1-m^*/\cohtime)$. The expressions \eqref{eq:r(d)_coh} and \eqref{eq:r(d)_noncoh} describe elegantly and succinctly the tradeoff between diversity gain and multiplexing gain at high SNR.

Note that $m^*/\cohtime$ is roughly the number of pilots per time-frequency slot needed to learn the channel at the receiver when $m^*$ transmit antennas are used. A comparison of \eqref{eq:r(d)_noncoh} with \eqref{eq:r(d)_coh} thus illustrates how an analysis of the diversity-multiplexing tradeoff under the assumption of no CSI at the receiver captures the channel-estimation overhead.

It has been recently demonstrated that for data packets of $1000$ channel uses or more and for moderately low packet-error probabilities (around $10^{-2}$), one should typically operate at maximum multiplexing \cite{lozano10-09a}. In this regime, which is relevant for current cellular systems, diversity-exploiting techniques are detrimental both for high- and for low-mobility users. 
For high-mobility users (where $\cohtime$ is significantly smaller than the packet size $n$), abundant time and frequency selectivity is available, so diversity-exploiting techniques are superfluous. For low-mobility users (where $\cohtime$ is large), the fading coefficients can be learnt at the transmitter and outage events can be avoided altogether by rate adaptation.

However, when the packet size becomes small and/or smaller packet-error probabilities are required, these conclusions may cease to be valid. For example, for packet lengths of, say, $100$ channel uses (which is roughly equal to a LTE resource block) and packet error probability of $10^{-5}$ or lower, spatial diversity may be more beneficial than spatial multiplexing. Furthermore, when the coherence interval $\cohtime$ is small, the cost of estimating the fading coefficients may be significant and must therefore be taken into consideration.

Studies based on capacity or outage capacity are inherently incapable of illuminating the entire \emph{diversity-multiplexing-channel-estimation tradeoff}. Indeed, recall that the capacity is defined as the largest rate at which reliable communication is feasible as the packet length tends to infinity. Specialized to the block-fading channel, capacity is typically studied by letting the number of time-frequency diversity branches $\tfdiv$ grow to infinity while holding the coherence interval $\cohtime$ fixed. For example, when $\cohtime>1$ and no CSI is available, the capacity is given by \cite{zheng02-02a}
\begin{equation}
\label{eq:C_BF}
C(\rho) = m^*\left(1-\frac{m^*}{\cohtime}\right)\log\rho + \mathcal{O}(1)
\end{equation}
where $\mathcal{O}(1)$ comprises error terms that are bounded in the SNR. Observe that \eqref{eq:C_BF} reflects the cost of estimating the fading matrix through $\cohtime$, but it hides away the effects of spatial diversity, since by letting $\tfdiv$ tend to infinity we achieve an infinite time-frequency diversity gain already through coding. Conversely, the definition of outage capacity is based on the assumption that the coherence interval $\cohtime$ grows to infinity while the number of diversity branches $\tfdiv$ is held fixed (cf.\ Section~\ref{sub:capacity-vs-outage} where we chose $\tfdiv=1$). For example, in the absence of CSI, the outage capacity is given by \cite{ozarow94-05a}
\begin{equation}
\label{eq:Cout}
C_{\epsilon}(\rho) = \sup\left\{R\colon \inf_{\mathsf{Q}^{\tfdiv}} P_{\text{out}}(R,\mathsf{Q}^{\tfdiv})\leq \epsilon\right\}
\end{equation}
where $P_{\text{out}}(R,\mathsf{Q}^{\tfdiv})$ denotes the outage probability
\begin{equation}
\label{eq:Pout}
P_{\text{out}}(R,\mathsf{Q}^{\tfdiv}) = \Prob\lefto[\frac{1}{\tfdiv}\sum_{k=1}^{\tfdiv} \log\det\left(\mathsf{I}+\mathbb{H}_k^{H}\mathsf{Q}_k \mathbb{H}_k\right)\leq R \right]
\end{equation}
{and where the infimum in \eqref{eq:Cout} is over all positive-definite $\txant\times\txant$ matrices $\{\mathsf{Q}_1,\ldots,\mathsf{Q}_{\tfdiv}\}=\mathsf{Q}^\ell$ whose traces satisfy $({1}/{\tfdiv}) \sum_{k=1}^{\tfdiv} \tr(\mathsf{Q}_k)\leq\rho$. 
In \eqref{eq:Pout}, the symbol $\mathsf{I}$ denotes the identity matrix, and $(\cdot)^H$ denotes Hermitian conjugation.
}
For $\tfdiv=1$, the outage probability \eqref{eq:Pout} specializes to \eqref{eq:Pout_1}. 
Observe that \eqref{eq:Cout} captures the effects of spatial and time-frequency diversity through the dimension of $\mathbb{H}_k$ ($\txant\times \rxant$) and {the value of} $\tfdiv$. 
However, as already mentioned at the end of Section~\ref{sub:capacity-vs-outage}, it hides away the cost of estimating the fading coefficient, since for an infinite coherence interval $\cohtime$ the channel can be estimated perfectly without a rate penalty.

To investigate the entire diversity-multiplexing-channel-estimation tradeoff for small packet lengths, bounds on $\maxrate(n,\epsilon)$ were presented in \cite{yang12-09a,ostman14-08b,durisi15-12a}. 
Here, we provide an example, taken from~\cite{durisi15-12a}, which illustrates the benefit of a nonasymptotic analysis of the diversity-multiplexing-channel-estimation tradeoff.
Specifically, we consider a scenario based on the 3GPP LTE standard~\cite{lozano10-09a} where the packet size is $n=168$ symbols, which corresponds to $14$ OFDM symbols, each consisting of $12$ tones. 
We set the SNR to $6$ dB and the packet error rate to $10^{-5}$, which corresponds to a URC scenario, and compute the bounds on the maximum coding rate obtained in~\cite{durisi15-12a} as a function of the coherence time $\cohtime$ or, equivalently, the number of diversity branches $\tfdiv$ (recall that $n=\tfdiv\cohtime$) for a $2\times 2$ MIMO system. 
\begin{figure}[t]
  \centering
    \includegraphics[width=.45\textwidth]{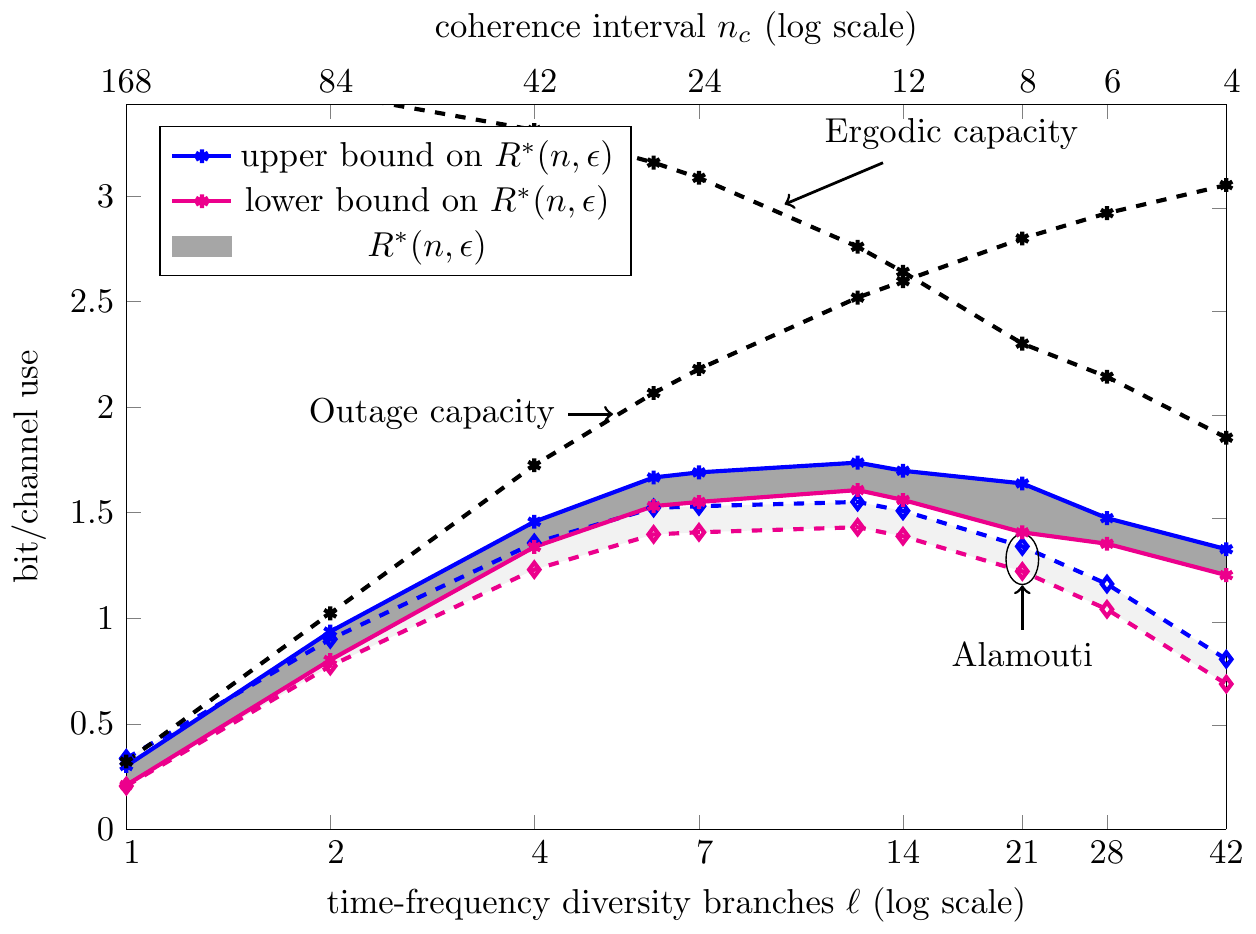}
  \caption{Upper and lower bounds on the maximum coding rate $\maxrate(n,\epsilon)$ for a Rayleigh block-fading channel with $\txant=\rxant=2$, $n=168$, $\epsilon=10^{-5}$, $\rho=6$ dB. The maximum coding rate lies in the shaded area between the upper and lower bound. Upper and lower bounds on the maximum coding rate achievable using an Alamouti inner code are also depicted to indicate the performance of a configuration in which transmit antennas are used to provide exclusively transmit diversity. The curve for the outage capacity has been obtained by numerically evaluating \eqref{eq:Cout}. The curve for the ergodic capacity follows by tightening \eqref{eq:C_BF}; see \cite{durisi15-12a} for more details. This figure appeared first in~\cite{durisi15-12a}.}
  \label{fig:figs_snr6eps05M2}
\end{figure}

The upper and lower bounds on the maximum coding rate obtained in~\cite{durisi15-12a} for the above example are depicted in Fig.~\ref{fig:figs_snr6eps05M2}.
We see from the figure that, given $n$ and $\epsilon$, the rate $\maxrate(n,\epsilon)$ is not monotonic in the coherence interval $\cohtime$, but there exists a value $\cohtime^*$ (in this case $14$) that maximizes the rate. 
This accentuates the fundamental tradeoff between time-frequency diversity (which decreases with $\cohtime$) and the ability of estimating the fading coefficient (which increases with $\cohtime$). 

We further observe that both outage capacity and capacity (computed for the scenario where CSI is not available at the receiver---see~\cite{devassy15-07a} for a recent review) fail to capture this tradeoff, although their intersection predicts surprisingly well the rate-maximizing coherence interval. Indeed, the outage capacity only captures the increase in time-frequency diversity, whereas capacity only captures the channel-estimation overhead.
We also note that when the coherence interval is smaller than $8$ channel uses, one of the two transmit antennas should be switched off, because the cost of estimating the fading coefficients overcomes the benefit of using two antennas at the transmitter.

In Fig.~\ref{fig:figs_snr6eps05M2}, we also depict bounds on the maximum coding rate obtainable using an Alamouti inner code~\cite{alamouti98-10a}, a configuration in which the transmit antennas are used to provide exclusively transmit diversity.
Since the gap between the rate achievable using Alamouti and the maximum coding rate converse is small, we conclude that for the scenario considered in Fig.~\ref{fig:figs_snr6eps05M2}, the available transmit antennas should be used to provide diversity and not multiplexing.

\subsection{Channel Dispersion versus Error Exponents}
Traditionally, the tradeoff between reliability and throughput for small packet lengths has been studied by means of \emph{error exponents}. In this section, we briefly discuss the relation between error exponents and asymptotic expansions of the maximum coding rate, such as \eqref{eq:approximation}, that express $\maxrate(n,\epsilon)$ as a function of channel capacity and channel dispersion.

Recall that the capacity $C$ is the largest transmission rate for which the packet error probability $P_{\text{e}}$ vanishes as the packet length $n$ tends to infinity. It turns out that for every fixed transmission rate $R<C$, the packet error probability vanishes even exponentially in $n$ \cite{feinstein55_09a}.  
It is therefore meaningful to expand $P_{\text{e}}$ for every fixed $R<C$ as
\begin{equation}
\label{eq:error_exponent}
P_{\text{e}} = e^{- n[E(R) + o(1)]}
\end{equation}
where $o(1)$ comprises remainder terms that vanish as $n$ tends to infinity. The exponent $E(R)$ in \eqref{eq:error_exponent} is referred to as the \emph{error exponent} corresponding to the rate $R$. For more details on error exponents, see \cite{gallager68a} and references therein.

Intuitively, \eqref{eq:error_exponent} characterizes the packet error probability $P_{\text{e}}$ as a function of $n$ and $R$. In contrast, \eqref{eq:approximation} characterizes the transmission rate $R$ as a function of $n$ and $P_{\text{e}}$. It may therefore seem plausible to view the expansions \eqref{eq:approximation} and \eqref{eq:error_exponent} as two equivalent characterizations of the triple $(R,n,P_{\text{e}})$. 
However,  \eqref{eq:approximation} and \eqref{eq:error_exponent} contain remainder terms, specifically $\mathcal{O}(\log n/n)$ and $o(1)$, and are therefore only accurate if the packet length $n$ is sufficiently large. 
Since $P_{\text{e}}$ decays exponentially in $n$, it follows that for packet lengths for which \eqref{eq:error_exponent} is a good approximation, $P_{\text{e}}$ is very small. 
Likewise, $\maxrate(n,\epsilon)$ converges to the capacity $C$ as $n$ tends to infinity, so for packet lengths for which \eqref{eq:approximation} is a good approximation, $\maxrate(n,\epsilon)$ is very close to $C$.

In summary, the error exponent $E(R)$ characterizes the triple $(R,n,P_{\text{e}})$ when the rate $R<C$ is held fixed and $P_{\text{e}}$ is very small. 
In contrast, the channel dispersion $V$ characterizes the triple $(R,n,P_{\text{e}})$ when $P_{\text{e}}\leq \epsilon$ is held fixed and $R$ is very close to capacity. 
For wireless communications, where a small but positive packet error probability can be tolerated, the asymptotic expansion of $\maxrate(n,\epsilon)$ provided in \eqref{eq:approximation} seems more meaningful.

\begin{figure*}[t]
  \centering
    \includegraphics[width=.75\textwidth]{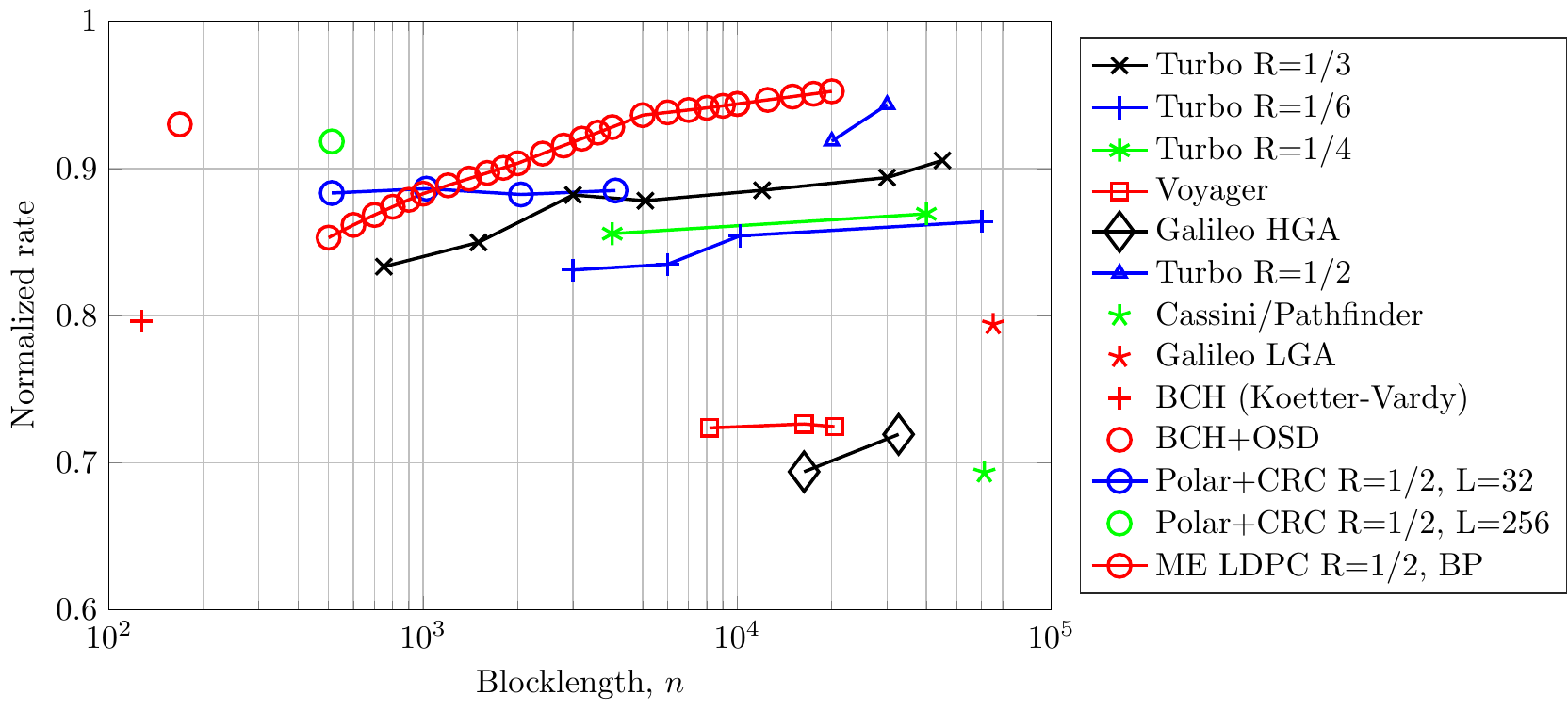}
  \caption{Normalized rates (with respect to $\maxrate(n,\epsilon)$) over binary-input AWGN channel; $\epsilon=10^{-4}$.
An earlier version of this figure appeared first in~\cite{polyanskiy10-11a}.}
  \label{fig:figs_universe}
\end{figure*}

\subsection{Further Works}
The work by Polyanskiy, Poor, and Verd\'u~\cite{polyanskiy10-05a} has triggered a renewed interest in the problem of finite-blocklength information theory. This is currently a very active research area.
Here, we provide a (necessarily not exhaustive) list of related works dealing with wireless communications at finite blocklength. 

When CSI is available at the receiver, the dispersion of  fading channels was obtained in \cite{polyanskiy11-08b,vituri12-11a,collins14-07a} for specific scenarios. 
Upper and lower bounds on the second-order coding rate of quasi-static multiple-input multiple-output (MIMO) Rayleigh-fading channels have been reported in \cite{hoydis13-03a} for the asymptotically ergodic setup when the number of antennas grows linearly with the blocklength. The channel dispersion of single-antenna, quasi-static fading channels with perfect CSI at both the transmitter and the receiver and a long-term power constraint has been given in \cite{yang15-07a,yang14-07a}.

For discrete-memoryless channels, feedback combined with variable-length coding has been shown to dramatically improve the speed at which the maximum coding rate approaches capacity~\cite{polyanskiy11-08a}. Such improvements can be achieved by letting the receiver feed back a stop signal to inform the transmitter that decoding has been successful (\emph{stop feedback}, also known as \emph{decision feedback}). 
One can relax the assumption that decoding is attempted after each symbol, with marginal performance losses~\cite{chen13-09a}.

Coding schemes approaching the performance predicted by finite-blocklength bounds have been also proposed. 
In~\cite{tal11-08a}, list decoding of polar codes is shown (through numerical simulations) to operate close to the maximum coding rate. The finite-blocklength gap to capacity exhibited by polar codes has been characterized up to second order (in terms of the so-called scaling exponent) in \cite{hassani14-10a,mondelli15-01a,goldin14-11a}. A comparison between the finite-blocklength performance of convolutional codes (both with Viterbi and with sequential decoding) and LDPC codes is provided in~\cite{maiya12-05a}.
Bounds and exact characterizations on the error-vs-delay tradeoff for codes of very small cardinality have been recently provided in~\cite{chen13-11a}.

In Fig.~\ref{fig:figs_universe}, we provide an overview of the performance of codes for the binary-input AWGN channel from 1980 to present.
The first eight codes in the legend of Fig.~\ref{fig:figs_universe} are from~\cite{dolinar98-a}.
The BCH (Koetter-Vardy) code is from~\cite[Fig.~2]{koetter01-05a}; here, the decoder uses soft-decision list decoding.
As shown in the figure, ordered-statistic decoding (OSD)~\cite{fossorier95-09a} of BCH codes improves the performance further. 
OSD decoding of nonbinary LDPC codes turn out to yield similar performance as BCH-OSD.
Indeed, this decoding technique seems to yield state-of-the-art performance for very short packets (between $100$ and $200$).
For larger packet size, list decoding of polar codes combined with CRC~\cite{tal15-05a} and multi-edge (ME) type LDPC codes~\cite{richardson04-a} are a competitive benchmark.

Moving to coding schemes exploiting decision feedback,  designs based on tail-biting convolutional codes combined with the reliability-output Viterbi algorithm have been proposed in~\cite{williamson14-10a}. 
%For short blocklengths, these schemes operate above the achievability bound provided in~\cite{polyanskiy11-08a}.
Finally, second-order characterizations of the coding rates for some problems in network information theory have recently been obtained. A comprehensive review is provided in~\cite{tan14-a}.

\subsection{\texttt{spectre}: short-packet communication toolbox} % (fold)
\label{sec:spectre_short_packet_toolbox_for_wireless_engineers}
%
%
%\begin{figure}[t]
%  \centering
%    \includegraphics[width=.15\textwidth]{./figs/spectre_lr.pdf}
%  \caption{A freely available collection of numerical routines for finite blocklength information-theoretic analyses.}
%  \label{fig:fbl-magazine_fbl-magazine-draft_figs_spectre}
%\end{figure}
%
%
To optimally design communication protocols for short-packet transmission, one needs to rely on accurate physical-layer performance metrics.
\texttt{spectre}--\emph{short-packet communication toolbox}~\cite{durisi14-12b} is a collection of numerical routines for the evaluation of upper and lower bounds on the maximum coding rate for popular channel models, including the AWGN channel, the quasi-static fading channel, and the Rayleigh block-fading channel.
This toolbox can be freely accessed online and is under development. 
%The participation of additional members of the information and communication theory communities to this endeavor is welcomed.  
All the numerical simulations reported in this paper can be reproduced using \texttt{spectre} routines.
% subsection spectre_short_packet_toolbox_for_wireless_engineers (end)
\section{Communication Protocols for Short Packets} % (fold)
\label{sec:rethinking_the_mac_layer}

%\todo{Todo Petar: Is there really no scientific work on protocol-design for short-packet transmission?}

In simple terms, a communication protocol is a distributed algorithm that determines the actions of the actors involved in the communication process. Protocol information, also referred to as metadata or control information, can be understood as a \emph{source code} \cite{gallager76-07a} that ensures correct operation of the protocols and describes, e.g., the current protocol state, the packet length, or the addresses of the involved actors.

Only few results are available on the information-theoretic design of communication protocols, e.g.,~\cite{EphremidesHajek,Laneman,WangAbouzeid}, and most of them deal with the (source coding) problem of how to encode the network/link state that needs to be communicated as a protocol information.
The problem of how to transmit the protocol-related metadata has been largely left to  heuristic approaches, such as the use of repetition coding. Broadly speaking, whereas information theorists busy themselves with developing capacity-approaching schemes for the reliable transmission of the information payload, they often see the design of metadata as something outside their competence area, or as stated in \cite{polyanskiy11-08a}: ``\dots control information is not under the purview of the physical layer \dots" Such a line of thinking is fully justifiable when the ratio between the data and metadata is the one depicted in Fig.~\ref{fig:figs_data_metadata}(a), where the metadata occupy a small fraction of the overall packet length. However, for applications where the data is comparable in size to the metadata---see  Fig.~\ref{fig:figs_data_metadata}(b)---this approach seems questionable.

In the following, we shall argue that a thorough understanding of how the maximum transmission rate $\maxrate(n,\epsilon)$ depends on the packet length $n$ and on the packet error probability $\epsilon$ is also beneficial for protocol design. As mentioned above, only few results are available on the information-theoretic design of protocols, and there is even less work that considers protocol design for short-packet transmission, e.g., \cite{makki15_06a,makki15_08a,fasarakis15_05a,alevizos15_06a}. This section is therefore based on three simple examples that illustrate how the tradeoffs brought by short-packet transmissions affect protocol design. We believe that these examples unveil a number of interesting tradeoffs worth exploring and we hope that they may motivate the research community to pursue a better theoretical understanding of protocol design.

For simplicity, we assume throughout this section an AWGN channel with SNR $\rho=10$, and we approximate $\maxrate(n,\epsilon)$ as
\begin{IEEEeqnarray}{rCL}\label{eq:approximation_examples}
  \maxrate(n,\epsilon) \approx C-\sqrt{\frac{V}{n}}Q^{-1}(\epsilon) + \frac{1}{2n} \log n
\end{IEEEeqnarray}
where $C$ and $V$ are given in \eqref{eq:C_Gauss} and \eqref{eq:V_Gauss}, respectively.\footnote{Recall that, as mentioned in Section~\ref{sec:rethinking_phy_performance_metrics}, replacing the remainder terms in \eqref{eq:approximation} by $\frac{1}{2n}\log n$ yields a good approximation for $\maxrate(n,\epsilon)$.} 
%It is expected that the tradeoffs encountered for the AWGN channel will also be relevant for fading channels. 
{We expect that tradeoffs similar as the ones we shall illustrate for the AWGN case will occur also for the fading case (see~\cite{durisi14-05a} for an example that supports this claim).
}
Solving \eqref{eq:approximation_examples} for $\epsilon$ yields the following approximation of the packet error probability as a function of the packet length $n$ and the number of information bits $k=Rn$ which we shall use throughout this section:
\begin{equation}
\label{eq:approximation_eps}
\epsilon^*(k,n) \approx  Q\lefto(\frac{nC - k+(\log n)/2}{\sqrt{nV}}\right).
\end{equation}

\subsection{Reliable Communication Between Two Nodes}
\begin{figure}[t]
  \centering
    \includegraphics{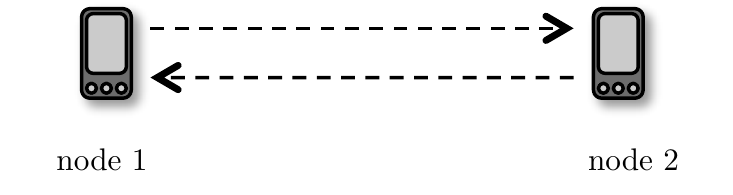}
  \caption{Scenario of a two-way communication with data from node 1 and acknowledgement from node 2.}
  \label{fig:2way}
\end{figure}
Consider the two-way communication protocol illustrated in Fig.~\ref{fig:2way}, where the nodes acknowledge the correct reception of a data packet by transmitting an ACK. The correct transmission of a data packet from, say, node 1 to node 2 would result in the following protocol exchange sequence:
\begin{enumerate}
\item The packet from node 1 is correctly received by node 2. We shall denote the probability of this event by $1-\epsilon_1$;
\item Node 2 sends an ACK to node 1. We shall denote the probability that an ACK is received correctly by $1-\epsilon_2$.
\end{enumerate}
As noted in \cite{bertsekas92-a}, if we communicate over a noisy channel and we are restricted to use a finite number of channel uses, then no protocol will be able to achieve perfectly reliable communication. Indeed, it is possible that either a packet is received incorrectly (an event which has probability $\epsilon_1$) or that the ACK is received incorrectly (which happens with probability $\epsilon_2$). By \eqref{eq:approximation_eps}, decoding errors are particularly relevant if the packet size is small, in which case $\epsilon_1$ and $\epsilon_2$ are large. Thus, the often-made assumptions of perfect error detection or perfect ACK-transmission (so-called ``1-bit feedback") are particularly misleading if the considered packet length is small.

Let us consider the following example. 
Let each node have a $6$-byte address and assume that node 1 has $12$ data bytes to send. Assume that the packet sent by node 1 contains the source address, the destination address, one bit for flow control and the data bytes. 
Hence, node 1 transmits $k_{i,1}=96$ data bits and $k_{o,1}=97$ metadata bits, resulting in $k_1=k_{i,1}+k_{o,1}=193$ bits. 
The ACK packet sent by node 2 consists of the source address and the destination address and one ACK bit.\footnote{Note that the source/destination addresses are necessary in order to uniquely identify the link to which the ACK belongs.} 
For the ACK packet, this yields $k_{i,2}=0$ data bits and $k_{o,2}=97$ metadata bits, so $k_2=k_{i,2}+k_{o,2}=97$ bits. 
Let $n$ be the total number of channel uses available to send the data and the ACK. 
{To optimize the protocol, we may want} to find the optimal number of channel uses $n_1$ by node 1 and $n_2=n-n_1$ by node 2  such that the reliability of the transmission, given by $\bigl(1-\epsilon^*(k_1,n_1)\bigr)\bigl(1-\epsilon^*(k_2,n_2)\bigr)$, is maximized. 
These values can be found numerically using the approximation \eqref{eq:approximation_eps}.
 For example, the minimum value of $n$ that offers reliability of transmission
\[\bigl(1-\epsilon^*(k_1,n_1)\bigr)\bigl(1-\epsilon^*(k_2,n_2)\bigr)>0.999\] is $n=203$, out of which $n_1=132$ channel uses are {for sending} the data packet and $n_2=71$ channel uses are {for sending} the ACK. 

As another example, fix $n=250$ as the maximal allowed number of channel uses. 
The numerical optimization that yields the largest reliability $\bigl(1-\epsilon^*(k_1,n_1)\bigr)\bigl(1-\epsilon^*(k_2,n_2)\bigr)$ gives $n_1=158$ and $n_2=92$. 
The resulting reliability is almost $1$ and the resulting throughput is  $\bigl(1-\epsilon^*(k_1,n_1)\bigr)\bigl(1-\epsilon^*(k_2,n_2)\bigr){k_{i,1}/{n}}=0.384$ bits/channel use. 

In many cases, it is not practical to have variable values for $n_1$ and $n_2${, and} a fixed time division duplex (TDD) structure in which $n_1=n_2$ is  preferred. 
In such a structure, there is no need of explicit ACK packets, since the acknowledgement is typically piggybacked on a data packet. In order to align this scenario with the last example, we assume that $n_1=n_2=125$, such that the acknowledgment for the packet arrives within $n=250$ channel uses from the start of the data transmission. 
A packet sent by nodes 1 and 2 contains $194$ bits, of which $96$ are data bits, $96$ are bits for addresses, $1$ bit is for flow control, and $1$ bit for the acknowledgment.
 Evaluating~\eqref{eq:approximation_eps} for these parameters gives $\epsilon^*(k_1,n_1)=\epsilon^*(k_2,n_2)=0.0118$. Observe that the reliability is markedly decreased, although the throughput is almost doubled to $0.759$ bits/channel use. 

These simple examples show that adjusting the packet length and the coding rate has the potential to yield high reliability. 
Note, however, that flexibility in the packet length necessarily implies that the receiver needs to acquire information about it.
This means that the protocol needs to reserve some bits within each packet for the metadata that describes the packet length. Our simple calculations have not accounted for this overhead. 

The use of a predefined slot length yields a robust system design, since no additional error is caused by the exchange of length-related metadata. This indicates that, in designing protocols that support ultra-high reliability,  a holistic approach is required that includes all elements of the protocol/metadata that are commonly assumed to be perfectly received. 

\subsection{Downlink Multi-User Communication}
\begin{figure}[t]
  \centering
    \includegraphics{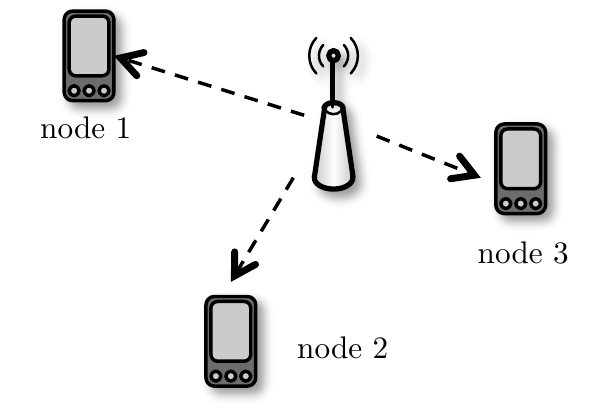}
  \caption{Example of a scenario with downlink communication from a Base Station over a broadcast channel to three nodes.}
  \label{fig:BC}
\end{figure}
We now turn to an example in which a base station (BS) transmits in the downlink to $M$ devices; see Fig.~\ref{fig:BC}. 
The BS needs to unicast $D$ bits to each device. 
Hence, it sends in total $MD$ bits. As a reference, we consider a protocol where the BS serves the users in a time division multiple access (TDMA) manner: each device receives its $D$ bits in a dedicated time slot that consists of $n$ channel uses. Thus, the TDMA frame consists of $M$ slots with a total of $Mn$ channel uses. 
In order to avoid transmission of metadata, we assume that the system operates in a \emph{circuit-switched} TDMA manner: (a) all devices and the BS are perfectly synchronized to a common clock; (b) each device knows the slot in which it will receive its data. 
The performance of this idealized scheme can be considered as an upper bound on the performance of practical systems, such as GSM, as it assumes that there is a genie that helps the devices remain synchronized.

The approximation on $\epsilon^*(k,n)$ in~\eqref{eq:approximation_eps} suggests that, for short packet sizes, it may be more efficient to encode a larger amount of data than the one intended to each device. 
Thus, instead of using TDMA, the BS may  \emph{concatenate} all the data packets for the individual devices. 
In this way, the BS constructs a single data packet of $MD$ bits that should be \emph{broadcasted} by using $Mn$ channel uses. Each receiving device then decodes the whole data packet and extracts the bits it is interested in from the decoded $M D$ bits. 

As a concrete example, assume that the BS wishes to transmit $D=192$ bits to each device and that there are $M=10$ devices. 
Furthermore, assume that $n=125$. 
We consider for simplicity one-shot communication. Accounting for retransmissions would require a more elaborate discussion.

In the reference scheme, the probability of error experienced by each device is $0.007$. If concatenation is used, however,  the probability of error drops to about $10^{-12}$, which  puts the transmission scheme in a different reliability class, while preserving the same overall delay. The price paid is the fact that each device needs to decode more data than in the reference scheme. 
%Also privacy and security considerations may make this second solution undesirable. 
 % However, this could be mitigated by considering a more elaborate scheme, in which the BS retransmits the erroneous packets, such that the total number of packets that need to be decoded in order to get the same chunk of data increases. \todo{Is it "increases" or "decreases"? It seems weird to have a more elaborate scheme that increases the number of packets to be decoded.}

Note that if one ignored the dependency of the packet error probability $\epsilon^*$ on the packet size $n$, one would conclude that the circuit-switched TDMA protocol is the most efficient, since all channel uses can be devoted to the transmission of payload bits. In contrast, by taking the dependence of $\epsilon^*$ on $n$ into account, we see that an unconventional protocol that concatenates the data intended to different devices outperforms the traditional TDMA protocol by orders of magnitude in terms of reliability. 

\subsection{Uplink Multi-User Communication}
\begin{figure}[t]
  \centering
    \includegraphics{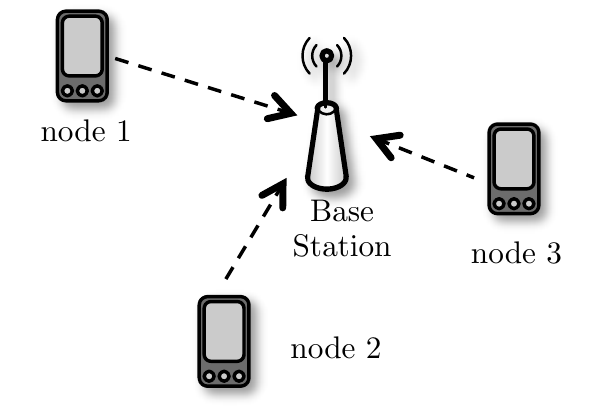}
  \caption{Example of a scenario with uplink communication from a set of three nodes over a multiple access channel to a common Base Station.}
  \label{fig:MAC}
\end{figure}
Our last example is related to the scenario depicted in Fig.~\ref{fig:MAC} in which $M$ devices run a random access protocol in order to transmit to a common receiver BS. 
Specifically, there are $M$ users, each sending $D$ bits to the BS. Each packet should be delivered within a time that corresponds to $n$ channel uses. 
These $n$ channel uses are divided into $K$ equally-sized slots of $n_K={n}/{K}$ channel uses. The devices apply a simple framed ALOHA protocol: each device picks randomly one of the $K$ slots in the frame and sends its packet. 
If two or more users pick the same slot, then a collision occurs and none of the packets is received correctly (see~\cite{durisi14-05a} for a more elaborate example). 
If only one device picks a particular slot (singleton slot), then the error probability is calculated using (\ref{eq:approximation_eps}) for $D$ bits and $n_K$ channel uses. 

We are interested in the following question: given $M$, $D$, and $n$, how should we choose the slot size $n_K$ in order to maximize the packet transmission reliability experienced by each individual device? 
This problem entails a tradeoff between the probability of collision and the number of channel uses available for each packet, which by~\eqref{eq:approximation_eps} affects the achievable packet error probability in a singleton slot.
Indeed, if $K$ increases, then the probability of a collision decreases, while the packet error probability for a singleton slot increases.
Conversely, if~$K$ decreases, then the probability of collision increases, while the packet error probability for a singleton slot decreases. 
The probability of successful transmission is given by
\begin{equation}\label{eq:ProbSuccessALOHA}
P_S=\frac{M}{K}\left( 1-\frac{1}{K}\right)^{M-1}\cdot  \bigl(1-\epsilon^*\left(D,n_K\right)\bigr).
\end{equation}
Here, $({M}/{K})\left( 1-{1}/{K}\right)^{M-1}$ is the probability of not experiencing collision, and $\epsilon^*\lefto(D,n_K\right)$ is the probability of error for a packet of $D$ bits sent over $n_K$ channel uses, which can be approximated by~\eqref{eq:approximation_eps}. 

As a concrete example, let us consider the setup where $D=192$ bits, $M=10$ devices, and $n=800$ channel uses. The number of slots that maximizes (\ref{eq:ProbSuccessALOHA}) is $K=6$.  In contrast, the classic framed-ALOHA analysis, which assumes that packets are decoded correctly if no collisions occur (i.e., $\epsilon^*=0$  in~\eqref{eq:ProbSuccessALOHA}), yields $K=M=10$. In fact, the same is true for any positive error probability $\epsilon^*$ that does not depend on $n_K$.

\section{Conclusions} % (fold)
\label{sec:conclusions}

Motivated by the advent of {novel} wireless applications such as massive machine-to-machine and ultra-reliable communications, we have provided a review of recent advances in the theory of short-packet communications and demonstrated through three examples how this theory can help designing novel efficient communication protocols that are suited to short-packet transmissions.
The key insight is that---when short packets are transmitted---it is crucial to take into account the communication resources that are invested in the transmission of metadata.
This unveils tradeoffs that are not well understood yet and that deserve further research, both on the theoretical and on the applied side.

\section*{Acknowledgment}
We would like to thank Yury Polyanskiy for letting us reproduce Fig.~\ref{fig:figs_universe}, Gianluigi Liva for bringing our attention to BCH codes combined with OSD (see Fig.~\ref{fig:figs_universe}), and Erik G. Str\"om for fruitful discussions. We further would like to thank the Managing Editor, Vaishali Damle, as well as the anonymous referees for their valuable comments.

% section  (end)
%%%%%%%%%%%%%%%%%%%%%%%%%%%%
 \bibliographystyle{IEEEtran}
 \bibliography{IEEEabrv,publishers,confs-jrnls,references}

\end{document}